# Can Large Language Models assist in Hazard Analysis?

Simon Diemert[1][0000-0001-9493-7969] and Jens H. Weber[1][0000-0003-4591-6728]

[1] University of Victoria, Victoria, Canada

**Abstract.** Large Language Models (LLMs), such as GPT-3, have demonstrated remarkable natural language processing and generation capabilities and have been applied to a variety tasks, such as source code generation. This paper explores the potential of integrating LLMs in the hazard analysis for safety-critical systems, a process which we refer to as *co-hazard analysis (CoHA)*. In CoHA, a human analyst interacts with an LLM via a context-aware chat session and uses the responses to support elicitation of possible hazard causes. In this experiment, we explore CoHA with three increasingly complex versions of a simple system, using Open AI's ChatGPT service. The quality of ChatGPT's responses were systematically assessed to determine the feasibility of CoHA given the current state of LLM technology. The results suggest that LLMs may be useful for supporting human analysts performing hazard analysis.

## 1 Introduction

So called Large Language Models (LLMs), such as GPT-3 [1] and BERT [2], have shown remarkable abilities in processing and generating text documents and images. Applications include machine translation, on-demand image generation, software development support (e.g., OpenAI's Codex model that powers GitHub Co-Pilot) [3–5]. The quality of text generated by LLMs is even high enough to cause legitimate academic integrity concerns in secondary and post-secondary settings [6, 7]. However, some authors have warned that, despite their ability to generate seemingly coherent text, LLMs do not understand the meaning of text and simply predict the next word/phrase using statistical models learned from very large training corpora [8–10].

Safety-critical systems often pose complex challenges for analysts tasked with their assurance. Many industrial standards provide guidance to practitioners in this area; for example, UL 4600 – *Safety of Autonomous Products,* ISO 26262 and ISO 21448 for automotive systems, EN 50126 for rail applications, and ARP4761 for civilian aircraft. *Hazard analysis* (HA) is a pivotal activity prescribed in these standards. At their core, HA methods provide a structured approach to reasoning, qualitatively or quantitatively, about how an event or condition (which we refer to as a 'hazard cause') can result in a loss event. HA methods involve some manner of brainstorming that relies on the judgement and creativity of analysts. This means that outcomes are sensitive to factors such as an analyst's level of experience, attentiveness, creativity, and knowledge of the system's underlying technical principles and relevant phenomena in its operating environment. Failure to perform an adequate HA can lead to increased risk arising from unknown (and unmitigated) hazard.

This paper proposes a co-operative method where a human analyst and an LLM both participate in HA. This is similar to the approach taken by Microsoft's GitHub Co-Pilot tool, which assists software developers by generating source code based on text prompts. However, instead of generating source code, our approach uses an LLM to suggest lines of inquiry for consideration by a human analyst thus providing a secondary source of ideas for the creative and error prone tasks related to HA. More specifically, the contribution of this paper is three-fold. First, a new co-operative hazard analysis (CoHA) process is introduced. Second, the feasibility, utility, and scalability of CoHA is studied in an experiment where an LLM (GPT-3 accessed through ChatGPT) is used to support an STPA for multiple versions of a water heater system.

## 2 Co-Hazard Analysis

HA is the process of *identifying*, *assessing*, and *controlling* hazards in safety-critical systems, where a hazard is defined as a potential condition of the system that, given worst case assumption about the state of the system or its operational environment, may contribute to the occurrence of a loss (accident or mishap) [11]. Several HA methods exist, including Failure Modes and Effects Analysis (FMEA), Hazard and Operability Study (HAZOP), System-Theoretic Process Analysis (STPA) [12, 13], and Functional Resonance Analysis Method (FRAM) [14].

HA is an iterative process across all phases in the system development lifecycle, reaching from conceptual design, via system construction up to operation, maintenance, and decommissioning. HA usually involves brainstorming and "what-if" discussions with participants from diverse perspectives. Prior experience with similar systems and historical data on their hazards and mishaps are commonly consulted.

The importance of HA for safety-critical systems development has motivated research focused on automating (aspects of) HA. Thomas used formal methods to automate HA using STPA and determine if an unsafe control action is adequately mitigated [15]. Millet *et al.* used formal modelling techniques to analyze the space of possible 'feature interactions' between system components to focus manual analysis on the most likely hazardous scenarios [16]. Several researchers have investigated HA for self-adaptive systems where a supervising system automatically analyzes proposed configurations of managed system to determine if a hazard can occur [17–20]. All these automated methods depend on a human analyst developing sophisticated system models that are amenable to computation via conventional algorithms. While the rigour demanded by these methods adds value (in terms of guiding critical thinking), it is also resource intensive and require skilled analysts to be successful.

Given recent advances in the field of artificial intelligence (AI), particularly LLMs, there has been an increasing interest in exploring the use of AI in system engineering processes. LLMs have already shown promise in assisting system engineering tasks [21]. We propose that conversational AI services like ChatGPT also have the potential to participate in hazard analysis as part of a human-AI team [22], for example during brainstorming sessions. We refer to this vision as Co-Hazard Analysis (CoHA).

Fig. 1 summarizes this preliminary vision, where human analysts provide the LLM with a preliminary description of the system (and related information, such as losses to be avoided) and then converse with it to identify hazardous conditions that may contribute to the occurrences of such losses. In Fig. 1 the 'conversation' with the LLM is depicted as "queries", perhaps of the form: "*Can {condition X} contribute to {hazard Y} occurring?*", "*How might a failure of {X} contribute to {hazard Y} occurring?*", or even "*Suppose {property Z}, can {condition X} still lead to {hazard Y}?*". The human analyst(s) review the responses of the LLM and then formulate the HA results.

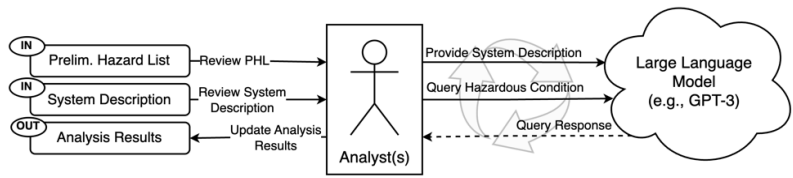

**Fig. 1.** Overview of CoHA method.

The role of the LLM is not to replace the human analysts. The AI is not even required to always provide "correct" answers. We believe that responses that may initially be seen as "incorrect" may in fact be useful to indicate problems with the system description, such as implicit (unstated) assumptions and ambiguities. However, involving the AI should provide, on average, add value to the HA process.

## 3   Experiment

This section explores the feasibility of CoHA with a preliminary experiment, focusing on the task of identifying unsafe control actions (UCAs) and causal (loss) scenarios as part of an STPA for a simple water heater system. The quality of the LLM's responses were assessed by two human analysts who are practiced with STPA.

STPA was chosen because: 1) it is a widely adopted hazard analysis method; 2) it includes an explicit system modelling step, which is helpful for structuring the input to the LLM; and 3) its UCA identification process uses explicit guide phrases (e.g., "not provided" or "stopped too soon") to assess how system behaviours might contribute to hazard occurrences, such phrases are readily used to build queries for an LLM. In this preliminary experiment we aim to answer the following research questions:

*RQ1 (Feasibility)* – *Can an LLM, when used for CoHA, produce results that are useful to human analysts identifying UCAs and causal scenarios for STPA?*

*RQ2 (Utility)* – *What proportion of responses from an LLM are useful and correct v. incorrect v. not useful for UCA and causal scenario identification?*

*RQ3 (Scalability)* – *Does the response quality of an LLM degrade as system under analysis increases in complexity?*

### 3.1 Method

To answer the research questions above, a four-phase experimental protocol was used: 1) system selection, 2) system encoding, 3) querying the LLM, and 4) manual response review and coding. This study used Open AI's ChatGPT platform (running GPT-3).

**System Selection.** The system for this experiment was chosen based on the following criteria. First, it should be expressible as a STAMP control structure, the modelling method for STPA. Second, it should be describable with simple terminology to avoid the case where the LLM's performance is degraded by niche terms or specialized concepts that might not appear in the training data for a general-purpose LLM. Third, the complexity should be scalable to assess performance on increasingly complex systems.

Per the criteria above, this experiment used a simply water heater with a simple feedback controller that aims to match the temperature of the water flowing out of a tank with a setpoint. In the simplest version of this system, water flows in and out of the tank and the controller provides a continuous 'enable' signal to a heating element that heats the water in the tank. A single thermometer provides temperature feedback to the controller. Complexity is increased by adding valves to control the flow in and out of the tank and by incorporating a secondary feedback loop within the physical ('plant') system, e.g., a heat exchanger. See Fig. 2 for depictions of these systems.

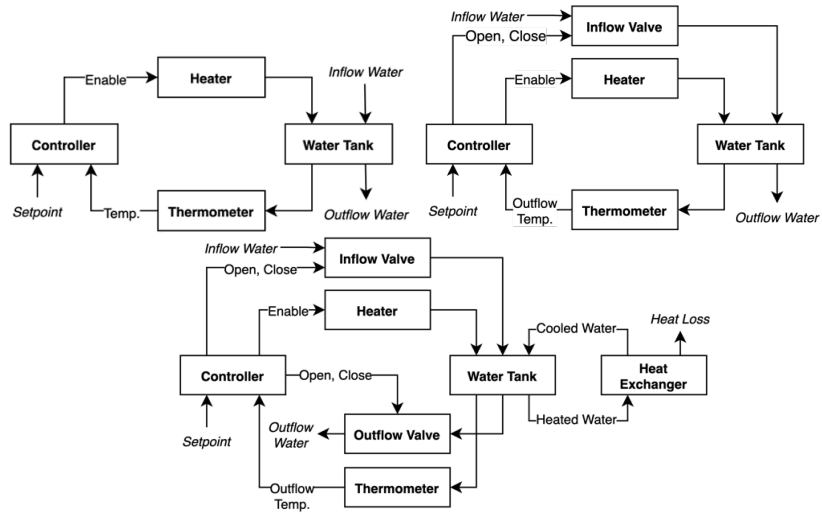

**Fig. 2.** Three STAMP control structures for the water tank system with increasing complexity: top-left – lowest, top-right – moderate, bottom-middle – highest.

**System Encoding.** Since ChatGPT allows a user to interact in a 'session' that stores context, the system description can be provided independently of and before the queries. A systematic approach was used to create a system description with four parts:

1. Elements - In an in-sentence list, name all system elements. Each block in the control structure should appear in this paragraph. For example: "*Consider a system consisting of a Controller, Heater, Water Tank, and Thermometer.*"

2. Relationships – Using short sentences, state the relationships between the elements listed in the first part. Every arrow between elements in the control structure should be represented in this paragraph. Template sentences are used to simplify statements and improve repeatability (e.g., "`{The|While|When|If} Element X provides Y [to Element Z]…`"). For the simple water heater, the description was:

   *The Controller provides the enable signal to the Heater to maintain a temperature setpoint. While the Controller is providing the enable signal to the Heater, the Heater heats the water in the Water Tank. When the Controller stops providing the enable signal to the Heater, the Heater does not heat the water in the Water Tank. The Thermometer measures the current water temperature inside the Water Tank. The Thermometer provides the current temperature of the water flowing out of the Water Tank to the Controller.*

3. Assumptions and Constraints – Statements describing additional constraints or assumptions are expressed in a third part. Short simple sentences are used like "`The Element X does not Y`" or "`Property W is never greater than V`". There are many constraints and assumptions that could be specified, even for a simple water heater system. For this experiment, some constraints and assumptions were intentionally not specified to allow us to observe how ChatGPT responds to under-specified systems. For the simple water heater, this part was:

   *The water flowing into the tank has variable temperature between 5 and 60 degrees Celsius. The ambient temperature is above 0 degrees Celsius. Water flows in and out of the tank at the same rate.*

4. Hazard Definition(s) – The definition of one or more hazards from the preliminary hazard list are provided in the last part. We adopt Leveson's definition of a hazard, i.e., a "system state or set of conditions that, together with a particular set of worst-case environment conditions, will lead to an accident (loss)" [13]. However, to avoid confusion with other definitions of 'hazard' that the LLM might have learned from the training data, the term "dangerous event" is used. Each dangerous event is expressed in a single sentence, for example: "*A dangerous event occurs if the temperature of the water flowing out of the tank is greater than 90 degrees Celsius.*"

Depending on the intent of the analysis, it is necessary to provide an additional statement about whether there are other unidentified dangerous events, i.e., does the analysis make a 'closed world` or 'open world' assumption. For this study, we adopted the closed world assumption and added the sentence: "*There are no more dangerous events*" to the final part of the system description. Preliminary experiments that omitted this sentence resulted in ChatGPT describing other potentially dangerous events for the water heater, like the water in the tank freezing.

**Querying the LLM.** Once the system description was provided to ChatGPT, queries were formulated using STPA's standard guidewords for identifying UCAs: "provided", "not provided", "too early", "too late", "stopped too soon", "applied too long". We omitted the "out of sequence" and "wrong order" guidewords that are also considered in STPA. Queries were formulated for each of the controller's control actions crossed with each identified dangerous event. Each query had the form: "`Could <Element X> <doing guideword + control action> result in <dangerous event>?`". For example: "*Could the Controller providing the enable signal too early to the Heater result in the temperature of the water flowing out of the tank exceeding 90 degrees C?*" Queries were provided one at a time to ChatGPT. For each query, the first response was recorded. Though the ChatGPT user interface provides options to pre-maturely halt response generation and re-generate responses, neither were used.

**Response Review and Coding.** The text of response was independently reviewed by two human analysts (the authors) who are both experienced users of the STPA method. Each word in the response (typically whole sentences) was coded into one of three categories based on how well the response would support a human analyst determining whether the candidate guideword-action pairing is relevant for hazard analysis.

- Correct and useful – The text is correct in that it: 1) accurately describes whether the guideword-action pairing correctly identifies a condition that precedes a hazard occurrence or correctly determines that a hazard does not occur; and 2) it does not contain any incorrect statements or invalid inferences. Text that provides a credible rationale to support a conclusion about hazard occurrences was also included.

- Correct but not useful – The text does not contain any incorrect statements or inferences but is not directly useful for a human analyst performing STPA. Such text might be regarded as 'noise' in the response. This includes text with repeated information (e.g., previously coded "useful and correct", or restated constraints or assumptions).

- Incorrect – The text contains incorrect statements, invalid inferences, and/or the response might be misleading to a human analyst.

Coding was first performed independently. Then the reviewers met to discuss their assigned codes. During discussion, reviewers were permitted to modify their codes to reflect an updated understanding of ChatGPT's response. Agreement between reviewers was measured after discussion using Cohen's Kappa agreement statistic [23]. Agreement was measured at the 'word level' such that complete agreement exists if both reviewers assigned the same code to each word. Agreement was computed for each response independently and again across the entire set of responses.

To answer the RQs, a 'final coding' was produced where each word in all responses is assigned a single code (i.e., a combination of the two reviewer's codes). In cases where both reviewers agreed, the agreed upon code was chosen. In cases of disagreement, the final code for the word was selected using the following protocol:

1. If one reviewer assigned the code "Correct and useful" and the other reviewer assigned the code "Correct but not useful", then the word was marked "Correct and useful" in the final coding. This is because 'usefulness' is a subjective, and it is possible that an arbitrary CoHA analyst would also find the word useful.
2. If one reviewer assigned the code "Correct but not useful" and the other reviewer assigned the code "Incorrect", then the word was marked as "Incorrect".
3. If one review assigned the code "Correct and useful" and the other reviewer assigned the code "Incorrect", the finding was indeterminate and the word was not included as part of the analysis used to answer the RQs.

Reviewers were also permitted to annotate ChatGPT's responses with additional observations. Themes that emerged from these comments are discussed below.

### 3.2 Results

In total, for all three system versions 78 queries were posed to ChatGPT. Additional summary statistics are shown in Table 1. The detailed results are presented in terms of the research questions posed above. The overall Cohen's kappa statistic was $\kappa = 0.89$, which represents significant (post-discussion) agreement between the reviewers.

Table 1. Results by system version (standard deviation in parentheses).

| Complexity | # Queries | Words per Response | Total Words | Agreement |
|---|---|---|---|---|
| Lowest | 6 | 143.7 (34.4) | 862 | 0.77 |
| Moderate | 28 | 122.7 (51.1) | 3436 | 0.91 |
| Highest | 44 | 116.0 (48.5) | 5016 | 0.89 |
| Overall | 78 | 120.6 (48.6) | 9404 | 0.89 |

**RQ1 – Feasibility.** For CoHA to be feasible, the LLM must return responses that (at a minimum) contain both correct and useful information for the human analyst. Even if only a portion of the response is correct and useful, it can still provide value to an analyst. Therefore, to answer RQ1 the number of responses that contain some information (at least one word) coded as 'Correct and useful' are considered. In total, across all system versions considered, 50 out of 78 (64%) of all responses contained some useful information (minimum number useful words was 23 words). See Table 2 for details.

Table 2. Number of responses with at least one word coded in each category.

| Complexity | Correct and Useful | Correct but not useful | Incorrect |
|---|---|---|---|
| Lowest | 5 (83%) | 5 (83%) | 3 (50%) |
| Moderate | 20 (71%) | 23 (82%) | 14 (50%) |
| Highest | 25 (56%) | 32 (72%) | 29 (70%) |
| Overall | 50 (64%) | 60 (77%) | 46 (59%) |

To further contextualize the results, an example query and response is shown in Fig. 3. The response contains samples of correct and useful (green), correct but not useful (blue), and incorrect (red) information. First, ChatGPT replies with a correct answer to the query. Then, it makes an incorrect statement about the relationship between the water temperature in that it neglects the role of inflowing water in cooling the tank's overall temperature; reviewers felt that the phrase '…primarily determined by…' was too strong. The next sentence has a mix of correct and incorrect information. While it is true that the inflow valve controls the flow rate into the tank, it is incorrect to say that it does not affect the temperature. The second last sentence is about the tank's water level and is not related to the query and is thus marked as 'correct but not useful'. Finally, the last two sentences offer observations about the system's behaviour, but they are not directly useful for answering the query.

> ***Query:*** *Could the Controller providing the open command to the Inflow Valve too early result in the temperature of the water flowing out of the tank exceeding 90 degrees C?*
>
> ***Response:*** *No, providing the open command to the Inflow Valve too early will not cause the temperature of the water flowing out of the tank to exceed 90 degrees C. The temperature of the water in the tank is primarily determined by the heater and the ambient temperature. The inflow valve controls the water flow rate into the tank, it does not affect the temperature of the water flowing out of the tank. The water level in the tank will increase due to the water flow rate. If the controller is monitoring the water level, it can adjust the inflow valve and the heater accordingly to maintain the temperature within the safe range.*

**Fig. 3.** Sample query and response from ChatGPT for highest complexity system. Key: green – correct and useful, blue – correct but not useful, red – incorrect.

Based on the above results, RQ1 is answered as follows:

> ***Answer to RQ1*** – *When used for CoHA with STPA, ChatGPT's responds to queries about potentially unsafe control actions with useful and correct information with moderate frequency (64% of the time). Therefore, it is feasible to use ChatGPT to support STPA. However, ChatGPT's responses are likely to also contain correct but not useful and incorrect information and so responses must be scrutinized by a human analyst.*

**RQ2 – Utility.** As shown in Table 1, responses from ChatGPT contain a mix of correct and useful, not useful, and incorrect information. For CoHA to be useful in broad sense, a human analyst must be able to easily use the LLM's responses. This is only possible if a meaningful proportion of each response is both useful and correct, and the proportions of not useful (i.e., 'noise') and incorrect information are relatively low. To answer RQ2 the proportion of words coded into each category is considered. Table 3 shows the distribution of codes in terms of total number and average number of words per response for each system version. Fig. 4 visualizes the overall proportion of words (across all responses for each system version) assigned to each code.

**Table 3.** Word-level codes applied to each system.

| Complexity | Correct and useful | | Correct but not useful | | Incorrect | |
|---|---|---|---|---|---|---|
| | Count | Avg (SD) | Count | Avg (SD) | Count | Avg (SD) |
| Lowest | 420 | 70.0 (39.0) | 343 | 57.2 (47.1) | 93 | 15.5 (19.7) |
| Moderate | 1461 | 52.2 (40.1) | 1296 | 46.3 (36.0) | 553 | 19.8 (24.4) |
| Highest | 1430 | 32.5 (32.5) | 1832 | 41.6 (46.1) | 1828 | 41.5 (42.3) |
| Overall | 3311 | 44.4 (37.3) | 3471 | 44.5 (42.5) | 2474 | 31.7 (36.9) |

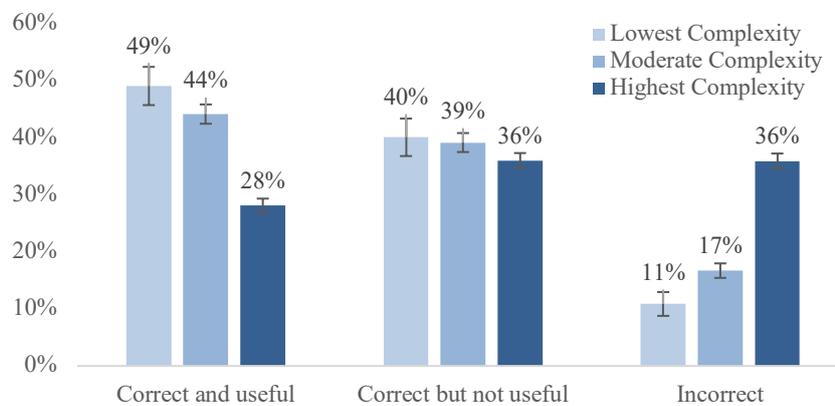

**Fig. 4.** Proportion of codes assigned by system complexity (95% CI shown).

It is observed that a meaningful proportion of ChatGPT's responses (especially for the lowest and moderate complexity systems) are correct and useful. Overall, across all system versions, 35% of words returned by ChatGPT were coded as 'correct and useful'. However, the proportion of correct and useful information appears to decline as the system complexity increases while the proportion of 'incorrect' information increases (see further analysis for RQ3). Interestingly, the amount of 'correct but not useful' information (overall 38%) is relatively consistent, even as the complexity increases.

Based on these results, RQ2 is answered as follows:

> ***Answer to RQ2*** – *When used for CoHA with STPA, between one quarter to one half of the content in ChatGPT's responses is correct and useful information. The remaining content is either correct but not useful to an analyst or simply incorrect. So, even though correct and useful information is available, human analysts performing CoHA with ChatGPT will have to sift through responses to find it. Therefore, in terms of correct and directly useful information, ChatGPT's responses are moderately useful.*

**RQ3 – Scalability.** As the complexity of the system increases, intuition says that the analysis should be harder (for either an AI or a human). RQ3 focuses on whether ChatGPT's responses degrade in quality as the complexity increases. Two word-level measures are of interest: 1) the proportion of incorrect words in ChatGPT's responses, and 2) amongst correct words in ChatGPT's responses, the proportion of useful words. Six pair-wise statistical (two-tailed) tests for significance between population proportions ($\alpha = 0.01$) for different system complexities are used [24]. Since multiple concurrent tests are performed, the Bonferroni correction is used to reduce the probability of a Type I error (i.e., incorrectly rejecting the null hypothesis) [25]. This resulted in a modified $a_0 = \alpha/6 = 0.00167$. The results are shown in Table 4.

As an example, consider the first row in Table 4, which corresponds to testing the null hypothesis, $H_0$, that the proportion of words coded as 'incorrect' is equal between the lowest and moderate complexity systems, $\hat{p}_{low}$ and $\hat{p}_{mod}$ respectively. From Table 3 above, the proportion of words coded as 'incorrect' for the lowest complexity system is computed as $\hat{p}_{low} = 93/(420 + 343 + 93) = 0.11$, similarly the proportion of 'incorrect' words for the moderate complexity system is $\hat{p}_{mod} = 0.17$. The difference between these proportions is $\hat{p}_{mod} - \hat{p}_{low} = 0.06$, which is a small increase in the proportion of incorrect words. The corresponding test statistic is $Z = 4.21$, which is larger than the critical value ($Z_{\alpha_0} = 3.14$). Therefore, the null hypothesis is rejected in favour of the alternative that the proportion of words coded as incorrect are different. The corresponding $p$-value is $p = 0.0000256$. From this test, it can be concluded that between the low and moderate complexity systems, the proportion of 'incorrect' words increased.

**Table 4.** Tests for significance between different system complexities.

| Measure | $H_0: \hat{p}_x = \hat{p}_y$ | $\hat{p}_x$ | $\hat{p}_y$ | $\hat{p}_y - \hat{p}_x$ | Outcome |
|---|---|---|---|---|---|
| Proportion of Incorrect Words in Responses | $\hat{p}_{low} = \hat{p}_{mod}$ | 0.11 | 0.17 | 0.06 | Reject $H_0$ ($p < 0.01$) |
| | $\hat{p}_{mod} = \hat{p}_{high}$ | 0.17 | 0.36 | 0.19 | Reject $H_0$ ($p < 0.01$) |
| | $\hat{p}_{low} = \hat{p}_{high}$ | 0.11 | 0.36 | 0.25 | Reject $H_0$ ($p < 0.01$) |
| Proportion of Useful Words in Responses | $\hat{p}_{low} = \hat{p}_{mod}$ | 0.55 | 0.53 | -0.02 | Do Not Reject $H_0$ |
| | $\hat{p}_{mod} = \hat{p}_{high}$ | 0.53 | 0.44 | -0.09 | Reject $H_0$ ($p < 0.01$) |
| | $\hat{p}_{low} = \hat{p}_{high}$ | 0.55 | 0.44 | -0.11 | Reject $H_0$ ($p < 0.01$) |

Inspecting the results in Table 4 shows that as the system complexity increases the proportion of words coded as incorrect increases. Additionally, amongst the correct words, the proportion that are useful words decreases with increased complexity, but only between the moderate and highest complexity systems (and therefore also the lowest to highest); the difference between the lowest and moderate complexity systems is inconclusive. Further quantification of the relationship between complexity and usefulness is not possible since we have not quantified the complexity of the systems.

Based on the above results, RQ2 is answered as follows:

> ***Answer to RQ3*** *– When used for CoHA with STPA, the quality of ChatGPT's responses declines as the system complexity increases. Specifically, for the example water heater systems, the proportion of incorrect information in ChatGPT's responses increases as system complexity increases. Additionally, for systems above a certain complexity, the proportion of correct and useful information declines.*

## 4   Summary and Discussion

This paper has introduced Co-Hazard Analysis (CoHA), a method where an LLM such as ChatGPT, assists a human analyst during HA. The results of our experiment allowed us to derive preliminary answers to our RQs. They indicate that CoHA, when used for STPA on simple systems, is a feasible and moderately useful method. However, it is also apparent that the performance of ChatGPT degrades as the system complexity is increased. Regardless, the results show that, with current LLM technology, CoHA has merit and warrants further investigation. The remainder of this section offers additional observations, discusses threats to validity, and proposes avenues for future work.

**Observations.** All queries posed during the experiment could reasonably be answered with a "yes" or "no" response; see the example in Fig. 3. However, ChatGPT's responses always contained a supporting rationale. In some cases, this rationale was factually incorrect or logically invalid, but in others it was both correct and concise. Indeed, even when it was incorrect, ChatGPT's rationale was more useful than just a binary assertion about whether a hazard might occur. We found that reviewing the rationale prompted us to think critically, often leading to further insights, even if those insights were not part of ChatGPT's response. Our experience suggests that the utility of CoHA might have been under-represented in our (empirical) answers to RQ2 and RQ3 above. Ultimately, we are interested in whether CoHA 'adds value' to the HA process, and so future studies might use methods that adopt a more wholistic notion of utility that includes the value of critical thinking prompted by an LLM's response.

For this initial study, the choice to issue binary queries, rather than more open-ended ones, was intentional, to simplify the review and analysis. However, a limitation of the binary queries is that they might have reduced ChatGPT's responses to a single causal scenario (in cases where it answered in the affirmative). Though sometimes ChatGPT replied with multiple scenarios to support its conclusion, this was not always the case. In reality, HA is an open-ended activity that would be difficult to accomplish through strict yes/no questioning. Therefore, follow-up experiments might explore the quality of an LLM's responses to more open-ended queries such as: "*Describe all ways that control action X results in dangerous event Y occurring?*".

Though the queries posed were binary in nature, we noted that ChatGPT sometimes 'hedges' and produces less-than-definitive responses. For example, instead of responding with "*No, X cannot happen*", it might reply with "*It is unlikely that X can happen*". Though we did not conduct a systematic analysis, ChatGPT appears to use "no" and "unlikely" (and similar terms) as synonyms, and it does not appear that it is able to grade the likelihood of specific events occurring so as to ascribe meaning to these words. Understanding whether LLM's are capable of 'fuzzy' reasoning in this manner would be an interesting line of future work.

During coding we noted that many responses from ChatGPT included suggestions for how to mitigate risk. For example, ChatGPT's responses often contained some variant of the sentence: *"... It's important for the controller to monitor the water temperature in the tank ..., and also to have safety mechanisms in place to prevent dangerous events..."*. While this is not an incorrect statement, it is also not useful in terms of answering the CoHA query about whether a specific condition could cause a hazard. Therefore, we marked these parts of the response as 'correct but not useful'. Regardless, it is interesting that ChatGPT is capable suggesting mitigations as part of its response. Using LLM's to synthesize mitigations might be a fruitful area of future research.

**Threats to Validity.** One threat to the study's internal validity arises from the inherent ambiguity of natural language, which is used to interface with ChatGPT. This might have impacted the study in three ways. First, the initial system description provided to ChatGPT was necessarily incomplete and did not include a specification of the system's behaviour for all operational scenarios. This was intentional since a human analyst can make reasonable assumptions about how a simple system will behave. As a trivial example, the system description in Section 3.1 does not explicitly state the response of the water's temperature to the heater being enabled (only that it "heats the water in the tank"). A human analyst would reasonably conclude that, all else equal, the temperature of the tank will increase when the heater is enabled. But it is not a given that a machine would make this inference on its own. The study method (partially) mitigated this concern, by adopting a template-based approach for describing the system to ChatGPT. Second, for similar reasons, there might have been ambiguity in the queries issued to ChatGPT. Different queries might produce different outcomes. Third, the responses from ChatGPT were reviewed by humans who have differing interpretations. Indeed, this occurred many times, most of which were resolved during discussion. In particular, we found that ChatGPT's responses often contained 'hedging' words that caused disagreement, e.g., the word "primarily" ChatGPT's response in Fig. 3.

Another threat to internal validity is the role of human reviewers. Though the study used a strict review protocol, it is possible that different reviewers would have arrived at a different final coding and thus different answers to the RQs. This compounds with the ambiguity in natural language if, for instance, a different set of reviewers would adopt a different consensus about how to interpret ambiguous words or phrases.

There are four notable threats to external validity. First, the study was performed using ChatGPT (Jan 9th version). It is possible that other LLM's or even other versions of ChatGPT will produce different results. Second, the study used STPA as an HA method, so we cannot directly say that CoHA is feasible or useful for other HA methods, such as FMEA or HAZOP. However, our overall impression on this front is positive: we hypothesize that, with appropriate inputs, CoHA will be feasible and useful for these HA methods as well. Third, the systems used for this study, even the highest complexity variant, are remarkably simple both in terms of the principles of their operation (i.e., basic operation of water heaters are broadly understood) and in terms of the number of components and interactions between them. It is not clear how ChatGPT would perform on a real-world system that is significantly more complex, though the results for RQ3 suggest its performance will be lower. Understanding the limits of LLM's for CoHA on complex systems remains an area of future work. Finally, the terminology used to describe the water heater systems is relatively simple ("tank", "heater", "valve", etc.) and likely well represented in the LLM's training data. Indeed, our experiment protocol included criteria to select systems that could be described using common language. However, real-world systems often contain niche concepts and terms that might be difficult or impossible to substitute. It is possible that LLM performance will degrade if more sophisticated terminology is required to describe systems.